\documentclass{elsart}
\usepackage[dvips]{graphicx}
\usepackage{xspace}

\newcommand{\SiH}{Si$_6$H$_6$\xspace}
\newcommand{\CH}{C$_6$H$_6$\xspace}

\advance\textwidth by 2 truecm
\advance\oddsidemargin by -1truecm

\advance\textheight by 0.5truecm

\begin{document}

\journal{Chemical Physics}
\volume{297}
\firstpage{13}
\lastpage{}
\pubyear{2004}
\runtitle{Stretched chemical bonds in \protect\SiH}
\runauthor{Grassi {\sl et al.}}

\begin{frontmatter}
\title{Stretched chemical bonds in \SiH:\\
A transition from ring currents to localized $\pi$-electrons?}

\author[CTChem]{A. Grassi,}
\author[CTChem]{G. M. Lombardo,}
\author[CTPhys]{R. Pucci,}
\author[CTPhys,corr]{G. G. N. Angilella,}
\author[Szeged]{F. Bartha,}
\and
\author[Oxford,RUCA]{N. H. March,}

\address[CTChem]{Dipartimento di Scienze Chimiche, Facolt\`a di
   Farmacia, Universit\`a di Catania,\\
Viale A. Doria, 6, I-95125 Catania, Italy}
\address[CTPhys]{Dipartimento di Fisica e Astronomia, Universit\`a di
   Catania,\\ 
and Istituto Nazionale per la Fisica della Materia, UdR di Catania,\\
Via S. Sofia, 64, I-95123 Catania, Italy}
\address[Szeged]{Department of Theoretical Physics, University of Szeged,\\
Tisza Lajos krt. 84-86, H-6720 Szeged, Hungary}
\address[Oxford]{Oxford University, Oxford, England}
\address[RUCA]{Department of Physics, University of Antwerp (RUCA),\\
   Groenenborgerlaan 171, B-2020 Antwerp, Belgium}
\thanks[corr]{Corresponding author.}

\begin{abstract}
Motivated by solid-state studies on the cleavage force in Si, and the
   consequent stretching of chemical bonds, we here study bond
   stretching in the, as yet unsynthesized, free space molecule \SiH.
We address the question as to whether substantial bond stretching (but 
   constrained to uniform scaling on all bonds) can result in a
   transition from ring current behaviour, characteristic say of
   benzene at its equilibrium geometry, to localized $\pi$-electrons
   on Si atoms.
Some calculations are also recorded on dissociation into 6 SiH
   radicals.
While the main studies have been carried out by unrestricted
   Hartree-Fock (HF) theory, the influence of electron correlation has 
   been examined using two forms of density functional theory.
Planar \SiH treated by HF is bound to be unstable, not all vibrational
   frequencies being real.
Some buckling is then allowed, which results in real frequencies and
   stability.
Evidence is then provided that the non-planar structure, as the Si--Si
   distance is increased, exhibits $\pi$-electron localization in the
   range 1.2--1.5 times the equilibrium distance.
\end{abstract}
\end{frontmatter}

\newpage

\setlength{\parskip}{0pt}
\setlength{\parindent}{2em}

\section{Introduction}

\noindent
In a fairly recent study, Matthai and March \cite{Matthai:97} have
   used a coordination-dependent force field constructed by Tersoff
   \cite{Tersoff:89} for crystalline Si, to calculate the cleavage
   force for two different cleavage planes.
This led them to propose that Si-Si bonds in the diamond lattice
   structure have comparable elasticity to the bond in the free space
   H$_2$ molecule.
Coulson and Fischer \cite{Coulson:49} made a
   variational calculation on H$_2$, using asymmetric orbitals centred on the
   two protons at separation $R$, which showed that conventional
   LCAO-MO was a viable approach until $R$ reached $1.6R_0$, with
   $R_0$ the equilibrium bond length in free space H$_2$.
The cleavage force calculations of Matthai and March \cite{Matthai:97} 
   led to a slightly larger value than $1.6$, characterizing the
   elasticity of the Si-Si bond in crystalline Si.

This has motivated us to study the stretching of Si-Si bonds in a
   different chemical environment, namely \SiH in free space.
Of course, in this, as yet unsynthesized, molecule, the natural
   `zero-order' hybrids are quasi-$sp^2$ in character, in contrast to $sp^3$ 
   in the diamond lattice structure of Si.
Therefore, the new feature in the bond stretching study undertaken
   here is the behaviour of the $\pi$-electrons, the MOs of which would
   conventionally be built, again in zero-order approximation, from
   linear combinations of $3p_z$ Si orbitals, the $z$ axis being
   perpendicular to the (assumed) planar molecule, with appropriate
   phase factors (see, \emph{e.g.,} Ref.~\cite{March:52a}).

It is important to say that strain energies of silicon rings and
   clusters have been studied earlier by Zhao and Gimarc
   \cite{Zhao:96}.
The work on bucky-tori by Babic \emph{et al.} \cite{Babic:01} is also
   noteworthy in the same context.
Finally, the relatively weak $\pi$-bonding ability of silicon versus
   carbon is emphasized in the study by Baldridge \emph{et al.}
   \cite{Baldridge:00}.

The outline of the paper is then as follows.
In Sec.~\ref{sec:2} immediately below, we record the results of a
   spin-compensated Hartree-Fock (HF) calculation of \SiH.
This predicts a stable molecule which is already `bound' at the HF
   level, with a substantial binding energy, even though no successful 
   synthesis has been achieved as yet, as already mentioned.

The spin-compensated HF treatment is then extended to the study of
   uniformly scaled \SiH, in which Si-Si and Si-H bonds are stretched
   in `benzene-like' geometry by a common scaling factor $\lambda$.
This spin-compensated HF treatment is then compared and contrasted with a 
   less restricted HF study in which, though a single determinantal
   wave function is still employed, different space orbitals are
   permitted for different spin directions.
Though constraints are still applied (\emph{e.g.,} the multiplicity is 
   inserted), such spin density states eventually prove to be lower
   than the spin-compensated HF energy as the scale factor is increased
   sufficiently.
Some details of the spin density are also referred to in
   Sec.~\ref{sec:4} and in the Appendix for the case of benzene.
Then, in Sec.~\ref{sec:new}, we have assessed the likely changes in the
   HF results due to electron correlation by employing density
   functional theory.
The major conclusions of the HF studies remain intact, the binding of
   the molecule with respect to SiH radicals being stabilized somewhat 
   by correlation inclusion.

\section{Equilibrium geometry and uniform stretching of Si-Si and Si-H 
   bonds by a scaling factor}
\label{sec:2}

\subsection{Equilibrium properties from spin-compensated Hartree-Fock
   (HF) calculations}
\label{ssec:2.1}

\noindent
In this section we shall first record the results of a
   spin-compensated (restricted) Hartree-Fock (RHF) calculation on the
   equilibrium geometry of \SiH.
We also anticipate results of analogous kind for \CH, to be referred
   to in the Appendix.
All calculations were performed with the {\tt Gaussian98} program
   \cite{Frisch:98} with both a 6-31G$^\ast$ and a cc-pVTZ
   basis set (for H and C atoms 
   \cite{Dunning:89}, and for Si atoms \cite{Woon:93}).
The total ground state energy $E_{\rm HF}$ of the single atoms and
   radicals (in unrestricted Hartree-Fock, UHF, approximation), and of
   the \SiH and \CH molecules (RHF) are recorded in Table~\ref{tab:1},
   together with the equilibrium bond lengths.
The binding energy relative to 6 Si and 6 H isolated atoms, as well as 
   that relative to 6 C and 6 H isolated atoms, is also recorded.

\subsection{Uniform stretching of all bonds by a common scale factor}
\label{ssec:2.2}

\noindent
With this as background, we turn to consider the stretching of
   chemical bonds in the \SiH molecule.
The stretching of Si-Si bonds was referred to already in connection
   with the cleavage force in crystalline Si in its diamond structure.
Of course, we are now considering such stretching in (a) a free space
   environment, and (b) in a situation in which $sp^2$ hybridization
   is replacing the basic $sp^3$ hybrids of the solid state (see also
   Stenhouse \emph{et al.} \cite{Stenhouse:77} for amorphous Si and
   its electron density distribution).
In this section, we shall confine ourselves to uniform stretching of
   all Si-Si and Si-H bonds in \SiH by a common scale factor $\lambda$.

\begin{figure}[ht]
\centering
\includegraphics[width=0.8\textwidth]{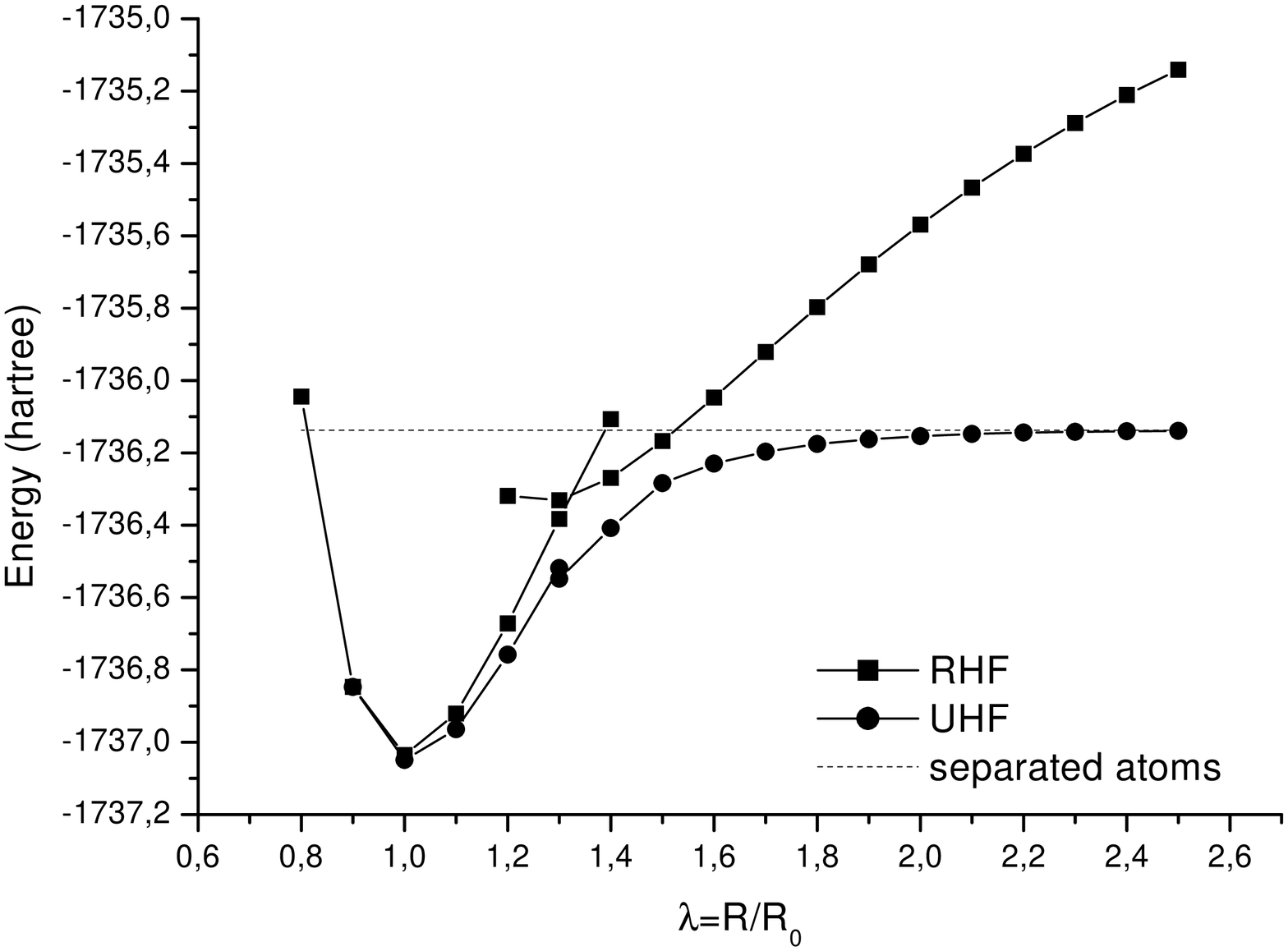}
\caption{Restricted (RHF) and unrestricted (UHF) Hartree-Fock
   potential energy curves of \protect\SiH as a function of uniform scaling
   factor $\lambda$.
}
\label{fig:1}
\end{figure}

Fig.~\ref{fig:1} shows the ground-state energy $E$ of \SiH in hartrees as a
   function of the uniform scaling factor $\lambda$ over a wide range
   of $\lambda$, covering both bond stretching ($\lambda>1$), and
   shortening ($\lambda<1$).
Equilibrium is achieved for $\lambda=1$ (no stretching), corresponding 
   to a `benzene-like' (planar) geometry, for which energies and bond
   lengths are given in Tab.~\ref{tab:1}. 
The dashed line in Fig.~\ref{fig:1} refers to the sum of atomic
   energies: \emph{i.e.,} that of 6 Si and 6 H isolated atoms.
Points marked by filled squares refer to a RHF calculation, whereas points marked
   by filled circles refer to an unrestricted Hartree-Fock (UHF)
   calculation, characterized by a state with a non-zero spin density, 
   built still on a single determinantal wave function, but now
   allowing different space orbitals for different spin directions.
All UHF curves here are with spin multiplicity $M=1$.
It is evident that the UHF result is always lower in energy than the
   RHF result, with the former tending to the correct dissociation limit
   (dashed line) as $\lambda$ increases.
It is found that the (putative) \SiH molecule binds for $\lambda \simeq 0.7 
   \div 1.3$, with a binding energy slightly lower than 1~hartree,
   measured relative to the dissociation limit.

Concerning the RHF curves, it turns out that as $\lambda$ increases,
   the symmetry properties of the solution change from those near
   equilibrium, which is the reason of the crossing of the RHF curves
   near $\lambda \sim 1.3$ in Fig.~\ref{fig:1}.
Specifically, while the symmetry of the HF state is a$_{1g}$ on both
   branches, the symmetry of the one-electron orbitals in the HF
   determinant is different.
Around equilibrium ($\lambda\simeq1$) we have [a$_{2u}$,e$_{1g}$] for the three
   highest occupied orbitals, which changes to [b$_{1u}$,e$_{2g}$] when bonds
   are further stretched ($\lambda>1.3$).

We shall return to the discussion of Fig.~\ref{fig:1}, when we
   deal with spin density in Sec.~\ref{sec:4} below.
However, we have also performed a variety of HF calculations in which
   only Si-Si bonds are uniformly stretched by $\lambda$, whereas Si-H 
   bonds are fixed at their equilibrium bond lengths.
The next section is devoted to comparing and contrasting the
   ground-state energies with those reported in the present section.

\section{Uniform scaling of Si-Si bonds only, with fixed Si-H bond
   lengths: Dissociation into 6 Si-H radicals}
\label{sec:3}

\begin{figure}[ht]
\centering
\includegraphics[width=0.8\textwidth]{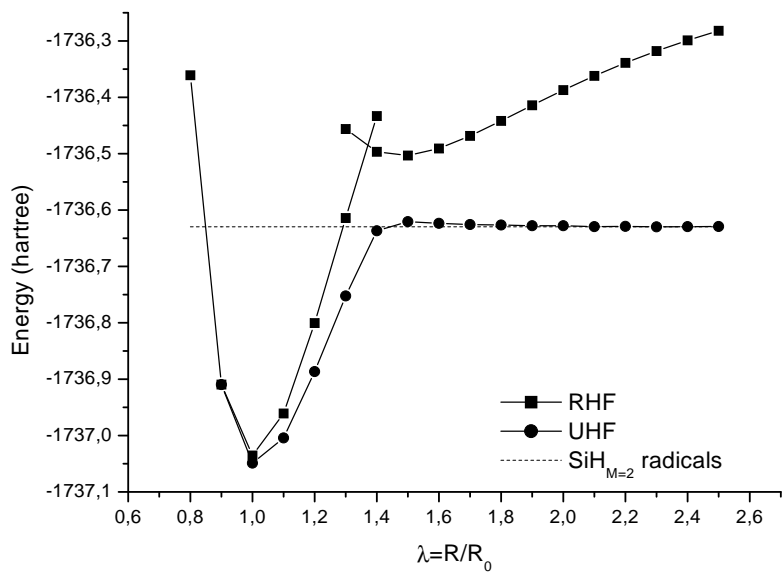}
\caption{Restricted and unrestricted Hartree-Fock
   potential energy curves of \protect\SiH.
Ring geometry of Si is scaled with $\lambda$, whereas Si-H bond
   lengths are kept fixed.
}
\label{fig:2}
\end{figure}

\noindent
In Fig.~\ref{fig:2} we show the RHF and UHF solutions, along with the
   dissociation limit (dashed line, corresponding to
   $E=-1736.6298$~hartree), in the case where the ring geometry of the 
   `benzene-like' structure of \SiH is uniformly stretched by a factor 
   $\lambda$, while the Si-H bond lengths are kept fixed at their
   equilibrium value.
Naturally, as $\lambda$ increases, the present case is energetically
   favoured with respect to the case of uniform stretching, because of 
   the gain in chemical binding energy due to the short Si-H bonds.

First, we focus again on the HF singlet case (RHF in
   Fig.~\ref{fig:2}).
For $\lambda\simeq 1$ (Fig.~\ref{fig:2}), the ground-state energy is
   approximately $-1737.05$~hartree.
The RHF solution deviates from the UHF one more markedly than in the
   uniformly stretched case, and crosses over a different symmetry
   branch well above the dissociation limit.
Following now the UHF curve, one observes that in this (albeit
   somewhat artificial) geometry, the system stays
   bounded only up to $\lambda\simeq 1.3$, namely for a smaller
   stretching than in the uniformly stretched case, whereas its energy 
   exceeds the dissociation limit at about $\lambda\sim 1.5$, where it 
   displays a weakly pronounced maximum (in the shape of an energy
   barrier), before decaying to the correct dissociation limit, as
   $\lambda$ is further increased.

\section{Itinerant {\em vs} localized $\pi$-electrons}
\label{sec:4}

\noindent
Since there is a sense in which UHF calculations reflect electron
   `correlation' by allowing different space orbitals for $\uparrow$
   and $\downarrow$ spins, it is relevant to enquire whether we can
   draw any conclusions about itinerant \emph{versus} localized
   behaviour of the $\pi$-electrons in \SiH.
A form for comparison is of course the `sister' molecule \CH.
In this latter case, the diamagnetic susceptibility and properties
   relating to ring currents testify, from experiment, to the
   delocalization of the $\pi$-electrons.

\begin{figure}[ht]
\centering
\includegraphics[width=0.8\textwidth]{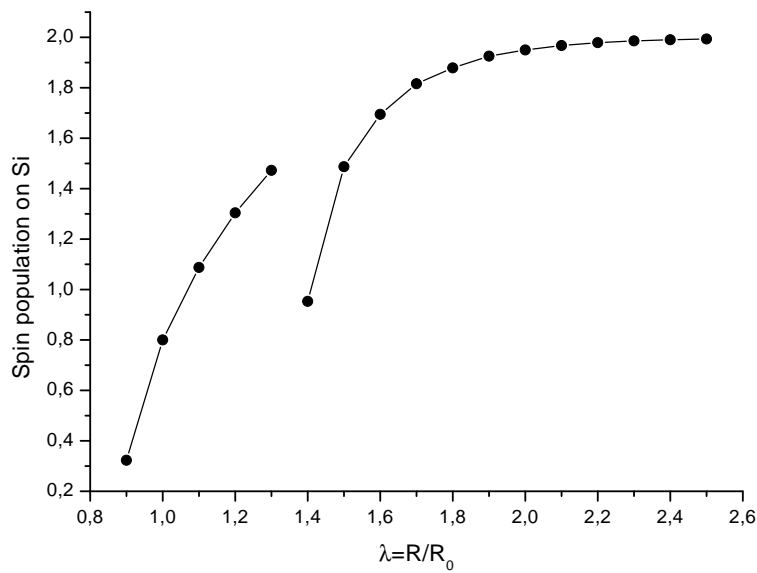}
\caption{Spin population magnitude on a Si atom in a \protect\SiH ring
   as a function of the uniform scaling factor $\lambda$.
Due to the symmetry of the UHF solution, spin populations have
   alternating signs on neighbouring Si atoms.
}
\label{fig:3}
\end{figure}

\begin{figure}[ht]
\centering
\includegraphics[width=0.8\textwidth]{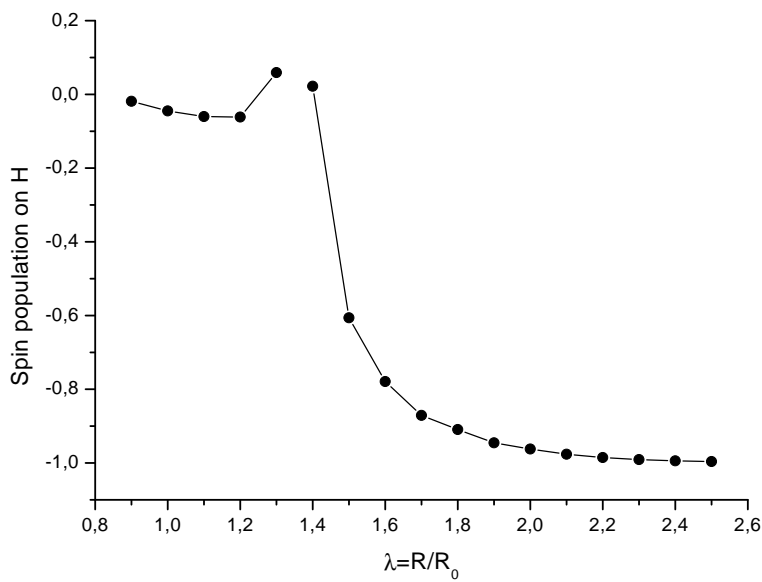}
\caption{Spin population on an H atom in a \protect\SiH ring as a function of
   the uniform scaling factor $\lambda$.
The sign is relative to that of the spin population on the Si atom
   belonging to the same radical (Fig.~\protect\ref{fig:3}).
}
\label{fig:4}
\end{figure}

In Figs.~\ref{fig:3} and \ref{fig:4} we show the spin population
   magnitudes on the Si and H atoms of the \SiH ring, respectively, as
   a function of uniform stretching $\lambda$.
Due to the threefold rotational symmetry of the lowest energy
   solution, spin populations had alternating signs on neighbouring
   radicals.

Within this approximation, $\pi$-electrons have their densities
   dominantly around the Si nuclei, with a tendency towards
   delocalization (ring currents) for low values of $\lambda$.

So far we have considered the structure of \SiH to be planar.
However, a study of the normal modes of vibration reveals that not all
   the vibrational frequencies are real.
This has led us to an investigation of the effect of buckling of the
   planar structure on these frequencies.

\begin{figure}[ht]
\centering
\includegraphics[width=0.8\textwidth]{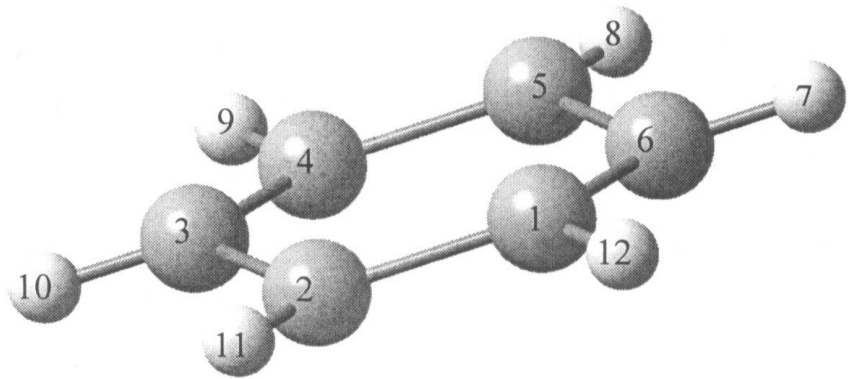}
\caption{
Optimized structure of \SiH.
See main text for discussion.
}
\label{fig:structure}
\end{figure}

The minimum energy structure has been obtained by means of an
   optimization procedure at the UHF-Singlet level using a
   6-31G$^\ast$ basis set.
The resulting molecular geometry is highly symmetric (D$_{3d}$ point
   group), similar to the cycloesane (C$_6$H$_{12}$) geometry in the
   chair conformation.
In particular, the Si--Si and Si--H bond lengths are $2.218$~\AA{} and
   $1.47$~\AA, respectively, while the valence angles are
   $119.28^\circ$ for Si--Si--Si and $119.31^\circ$ for Si--Si--H.
In contrast to planar 6-membered rings, in this structure only four atoms
   belong to a plane, say Si$_1$--Si$_2$--Si$_4$--Si$_5$, while the
   other two atoms, say Si$_3$ and Si$_6$, are one below and one above
   this plane (Fig.~\ref{fig:structure}).
Correspondingly, the angle between the four atoms plane with the
   Si$_2$--Si$_3$--Si$_4$ plane is $14.57^\circ$, while the angle
   between the four atoms plane and the Si$_1$--Si$_6$--Si$_5$ plane
   is $-14.57^\circ$.

The vibrational analysis confirmed that this structure corresponds to
   a local minimum, with all the 30 normal modes being real.
However, because of the symmetry, 10 frequencies are doubly
   degenerate, with a total of 20 single levels.
The infrared spectrum should be quite simple, with only 3 active
   modes, namely, the doubly degenerate level at $2407.9$~cm$^{-1}$,
   corresponding to the asymmetric stretching of the Si--H bond, the
   $800.5$~cm$^{-1}$ level, corresponding to the asymmetric bending of
   the Si--Si--H valence angle, and the $489.9$~cm$^{-1}$ level,
   corresponding to a composite motion of an \emph{out of plane}
   vibration of the H atoms and of an asymmetric stretching of the
   Si--Si bonds.

\begin{figure}[ht]
\centering
\includegraphics[width=0.8\textwidth]{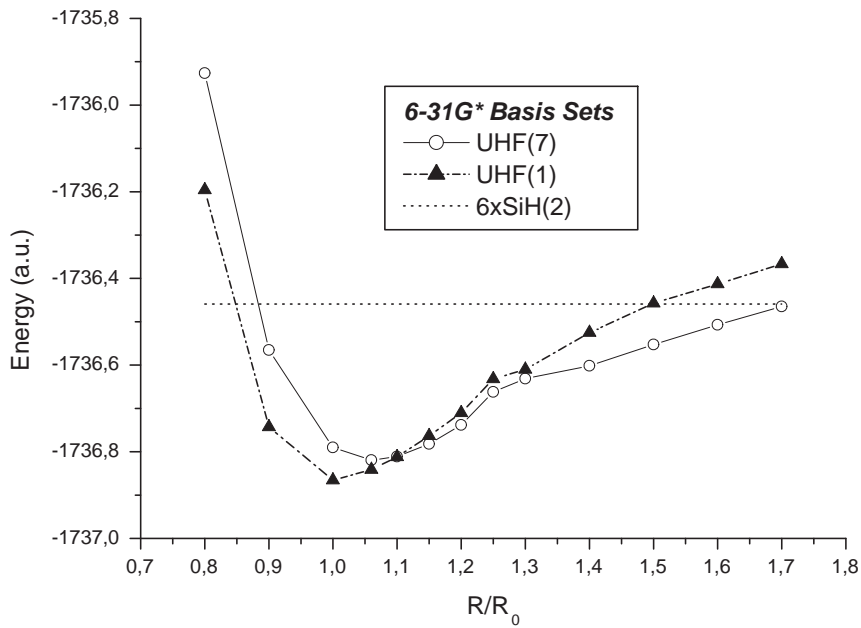}
\caption{
Unrestricted Hartree-Fock (UHF) potential energy curves of
   \protect\SiH, as a function of the uniform scaling factor
   $\lambda=R/R_0$, for the 6-31G$^\ast$ basis set.
See main text for discussion.
}
\label{fig:5}
\end{figure}

Calculations of electronic energies have then been performed which
   lead to Fig.~\ref{fig:5}.
This will now be discussed in some detail.
The 6-31G$^\ast$ basis set was employed within the UHF framework.

Starting from the fully optimized structure for the UHF-singlet (a
   structure very similar to that of the cyclo-hexane, with a
   ``chair''-like shape), the uniform scaling factor $R/R_0$ was
   increased from $0.8$ to $1.7$.
At each step, only the Si--Si distance and one internal torsional
   angle were kept constant.
This was required in order to allow a closed ring structure.
All the other degrees of freedom (including those of the hydrogen
   atoms) were optimized by the calculation.
The fully optimized UHF-singlet strcture has been used also for the
   UHF-7 calculations, for which the same procedure was employed:
   fixed Si--Si distance plus one internal angle, all the rest was
   left free to be optimized, including the hydrogen variables.
As can be seen from Fig.~\ref{fig:5}, up to $R/R_0 = 1.15$, the UHF-1
   is more stable than the UHF-7.
Such a situation corresponds to having the electrons localized on each
   Si $p$-orbital.
From $R/R_0 = 1.15$ on, the UHF-7 structures are more stable than the
   corresponding UHF-1, and the energy (correctly) tends to that of
   6~SiH radicals, each in a doublet configuration.

The conclusion is that, in this basis, the $\pi$ system localizes at a
   distance equal to 1.2--1.5 times the equilibrium distance.
As far as the spin population is concerned, in all the UHF-1
   calculations, the Gaussian program assigns the value 0 to every
   atom in the molecule.
On the other hand, positive values, with moduli
   about unity, are found at the end of each calculation for the UHF-7
   configurations.

\section{Summary and future directions}
\label{sec:5}

\noindent
We have here presented a variety of HF calculations for stretched
   chemical bonds in the, as yet unsynthesized, molecule \SiH.
The UHF state is lower in energy than the HF singlet state with
   uniform scaling of all bonds by a factor $\lambda$, throughout the
   range of $\lambda$ investigated in the present work ($\lambda
   \simeq 0.8$ to $2.6$), and a non zero spin density is predicted on
   Si (and H) nuclei, with alternating sign on neighbouring radicals.

In Sec.~\ref{sec:4} above, we have shown that, in contrast of course
   to benzene, the stable form predicted for \SiH at the present level
   of approximation is a non-planar structure.
The motivation for this was that the planar structure we initially
   investigated was `unstable' in the sense that not all the
   vibrational frequencies were real.
The `buckled' structure reported in Sec.~\ref{sec:4} has all its
   vibrational frequencies real.

Also, based on Fig.~\ref{fig:5}, we have presented evidence that for
   such a non-planar structure the $\pi$-electron assembly exhibits a
   tendency to pass from delocalized character at the equilibrium
   Si--Si length to localized behaviour when the bonds are stretched,
   the effect securing at a distance equal to 1.2 to 1.5 times the
   equilibrium distance.

Of course, for the future, it remains of considerable interest to see
   whether \SiH can be successfully synthesized.
From solid-state facts, however, Si seems not to `like' $sp^2$
   hybridization, there being only diamond structure in the
   crystalline state, with no analogue of $sp^2$ bonded graphite
   layers with C-bonding.
Nevertheless, the binding energy we have predicted relative to 6 SiH
   radicals is substantial
   ($\sim \frac{1}{2}$~hartree) and we could anticipate, from our
   earlier arguments 
   \cite{Grassi:96} relating electron correlation in bonds to bond
   order, that electron correlation would increase this binding
   energy.
This has been supported by the density functional results of
   Sec.~\ref{sec:new}.


\section*{Appendix A: Comparison with benzene with stretched bonds}

\noindent
To compare and contrast with the calculations reported in the main
   text, we summarize here results on benzene (\CH) with stretched bonds.
Fig.~\ref{fig:A1} shows the ground state energy as a function of
   uniform scale factor $\lambda$.
Clearly, the spin density HF solution rapidly becomes lower in energy
   than the RHF curve.

The evaluated spin densities on a C atom and on an H atom are shown in 
   Fig.~\ref{fig:A2}.
One presumes that, quite rapidly, there will be a transition in
   benzene from ring currents in the $\pi$-system to localized $\pi$
   electrons.
Fig.~\ref{fig:A3} shows the variation of the average value of the spin 
   operator ${\bf S}^2$ with stretching parameter $\lambda$.

Some calculations have also been performed with dissociation into 6
   independent CH radicals.
The results resemble those reported in Section~\ref{sec:3} of the
   main text for \SiH.
However, the UHF curve on Fig.~\ref{fig:A4} is always below the
   dissociation limit (compare with Fig.~\ref{fig:2}), and the spin
   population on a CH independent radical is 3 parallel spins (see
   Fig.~\ref{fig:A5}), whereas for a SiH radical one net spin is found
   to be more stable.

\section*{Appendix B: Approximate inclusion of electron correlation: density
   functional results}
\label{sec:new}

Having established the predictions of the unrestricted Hartree-Fock
   theory, among which the most significant for comparison with
   density functional predictions reported below are \emph{(i)} a
   Si--Si bond length in $\mathrm{Si_6 H_6}$ at equilibrium of
   $2.214$~\AA{} and \emph{(ii)} a binding energy of $0.40$~a.u. with
   respect to SiH radicals, we turn to density functional results as
   a means to make an approximate assessment of the effect of
   electron-electron Coulombic correlations.

The BLYP (Becke \cite{Becke:88}, Lee-Yang-Parr
   \cite{Lee:88}) and the  B3LYP (Becke
   \cite{Becke:93}, Lee-Yang-Parr \cite{Lee:88}) density functional
   calculations are reported in Table~\ref{tab:new}.
In contrast to the Hartree-Fock calculations, where the optimized SiH
   bond length for the ring was used throughout, the Si--H distance
   has now been optimized separately for radicals as well as for the
   ring.
The Si--Si bond length was optimized for the ring while it was
   constrained to a planar hexagon geometry.

For B3LYP, the Si--Si equilibrium bond length was found to be very
   close to the HF value of $2.214$~\AA{}, namely $2.215$~\AA.
For the BLYP functional, the bond was somewhat longer (by $0.02$~\AA).
The shorter bond length given by the B3LYP functional
   led to a somewhat greater binding energy $\Delta E$ with
   respect to the SiH radicals (bound with respect to atomization).
However, both functionals are embraced by $\Delta E = 0.547\pm
   0.005$~a.u., which exceeds the HF prediction by approximately 4~eV.
Correlation inclusion therefore does not change the essence of the HF 
   results for the molecule \SiH.

\begin{ack}
\noindent
N.H.M. wishes to acknowledge that his contribution to this study was
   made during several visits to the University of Catania.
It is a pleasure to thank Professor R. Pucci and his colleagues for
   generous hospitality.
N.~H.~M. also wishes to thank Professor V.~E. Van Doren for his
   continuing interest and support.
We also acknowledge very helpful discussions with Dr. F. Bog\'ar.
Two referees made penetrating comments on the first version of this
   paper, which led us to investigate non-planar structures.
Also we now have more convincing evidence for
   delocalization-localization transition in the $\pi$-electron
   assembly in \SiH as a result of their comments.
\end{ack}

\bibliography{Angilella,a,b,c,d,e,f,g,h,i,j,k,l,m,n,o,p,q,r,s,t,u,v,w,x,y,z,zzproceedings}
\bibliographystyle{prsty} 

\newpage

\bigskip
\begin{table}[ht]
\caption{Data for equilibrium configuration (uniform stretching
   parameter $\lambda=1$) of isolated Si, H, and C atoms, of SiH and
   CH isolated radicals, and of both \protect\SiH and \protect\CH
   molecules.
$M$ is the spin multiplicity; $E_{\rm HF}$ is the Hartree-Fock ground
   state energy.
The total energy relative to 6 Si and 6 H isolated atoms equals
   $E_{\rm HF}^0 (\mbox{6 Si + 6 H}) = -1736.13726$, whereas that for
   6 SiH isolated radicals equals $E_{\rm HF}^0 (\mbox{6 SiH}) =
   -1736.6298$.
Analogously, $E_{\rm HF}^0 (\mbox{6 C + 6 H}) = -229.14828$ and
   $E_{\rm HF}^0 (\mbox{6 CH}) = -229.73574$.
All energies and lengths are in hartrees and angstroms, respectively.}
\label{tab:1}
\medskip
\centering
\begin{tabular}{llrr@{.}lr@{.}lr@{.}l}
\hline
 & & $M$ & 
\multicolumn{2}{c}{$E_{\rm HF}$} &
\multicolumn{2}{c}{Spin population} &
\multicolumn{2}{c}{Bond length} \\
\hline
\hline
UHF & Si & 3 & $-288$ & 8564 &
\multicolumn{2}{c}{2} & \multicolumn{2}{c}{} \\
UHF & SiH & 2 & $-289$ & 4383 & 1 & 025 (Si) & 
   1 & 4727 (Si-H) \\
\multicolumn{5}{c}{} & $-0$ & 025 (H) & \multicolumn{2}{c}{} \\
RHF & \protect\SiH & 1 & $-1737$ & 035 &
   \multicolumn{2}{c}{} & 1 & 4727 (Si-H)  \\
\multicolumn{7}{c}{} & 2 & 2138 (Si-Si) \\
\hline
UHF & C & 3 & $-37$ & 69157 &
\multicolumn{2}{c}{2} & \multicolumn{2}{c}{} \\
UHF & H & 2 & $-0$ & 499810 &
\multicolumn{2}{c}{1} & \multicolumn{2}{c}{} \\
  & CH & 4 & $-38$ & 28929 & 2 & 940 (C) & 
   1 & 0734 (C-H) \\
\multicolumn{5}{c}{} & 0 & 060 (H) & \multicolumn{2}{c}{} \\
RHF & \protect\CH & 1 & $-230$ & 7805 &
   \multicolumn{2}{c}{} & 1 & 0734 (C-H)  \\
\multicolumn{7}{c}{} & 1 & 3827 (C-C) \\
\hline
\hline
\end{tabular}
\end{table}

\setcounter{figure}{0}

\renewcommand{\thefigure}{A.\arabic{figure}}

\begin{figure}[ht]
\centering
\includegraphics[width=0.8\textwidth]{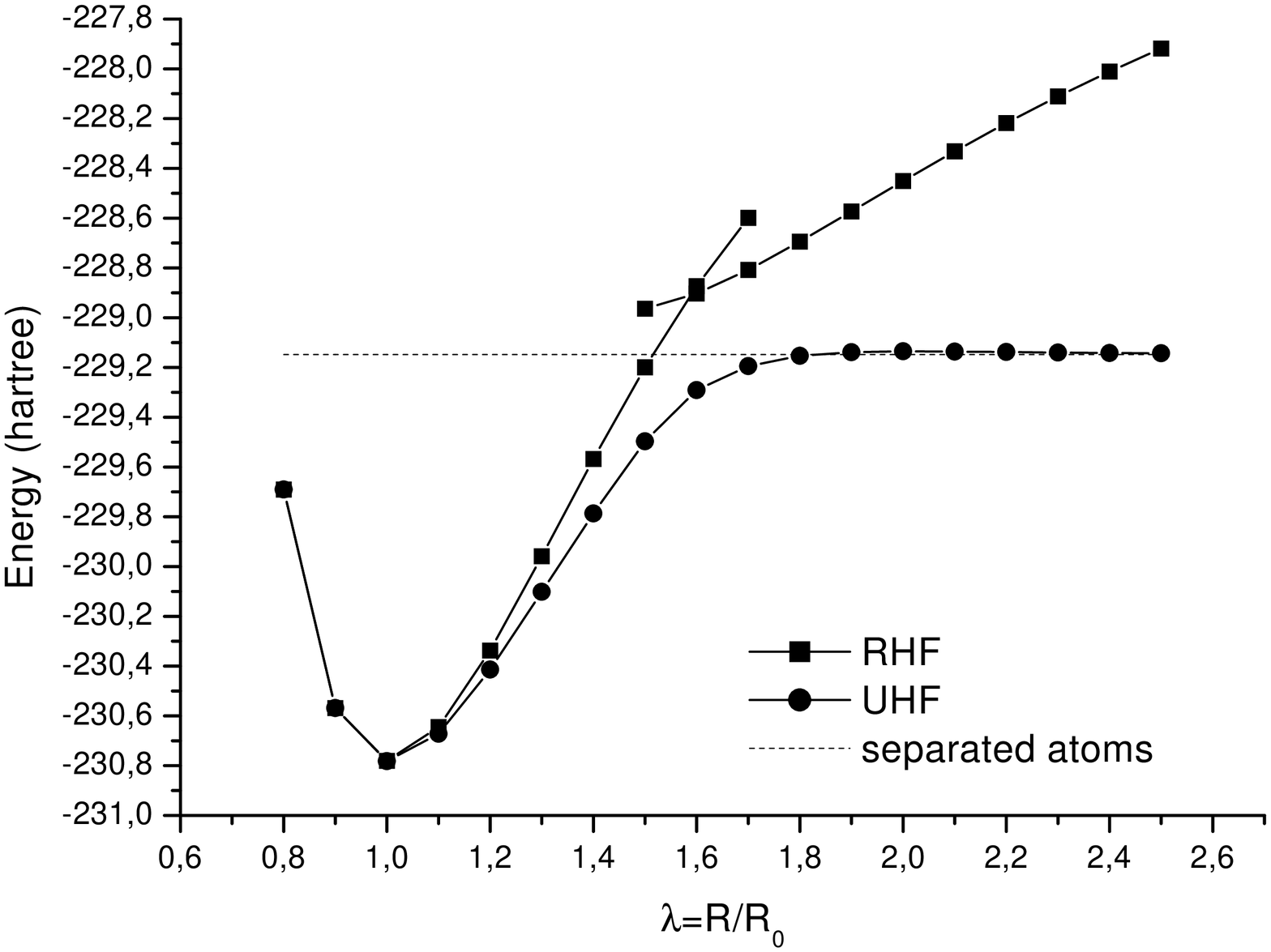}
\caption{Restricted and unrestricted Hartree-Fock
   potential energy curves of \protect\CH as a function of uniform scaling
   factor $\lambda$.
}
\label{fig:A1}
\end{figure}

\begin{figure}[ht]
\centering
\includegraphics[width=0.8\textwidth]{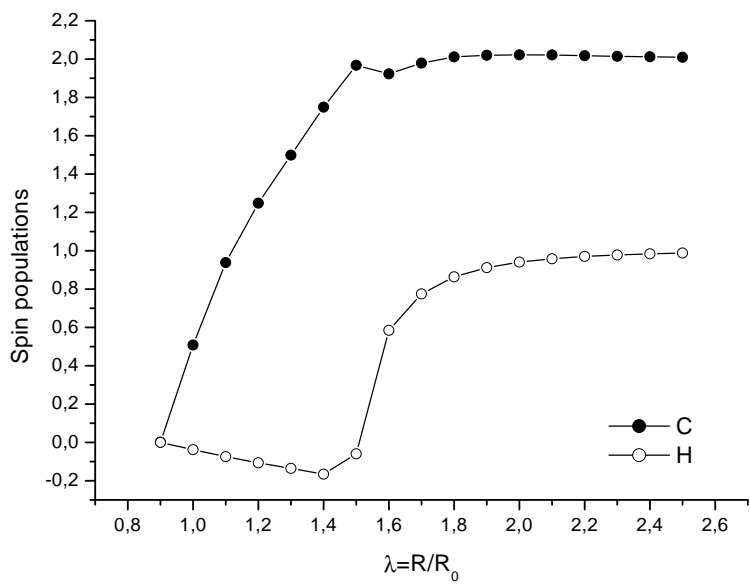}
\caption{Spin population on a C atom and the corresponding H atoms in
   a \protect\CH ring as a function of the uniform scaling factor $\lambda$.
}
\label{fig:A2}
\end{figure}

\begin{figure}[ht]
\centering
\includegraphics[width=0.8\textwidth]{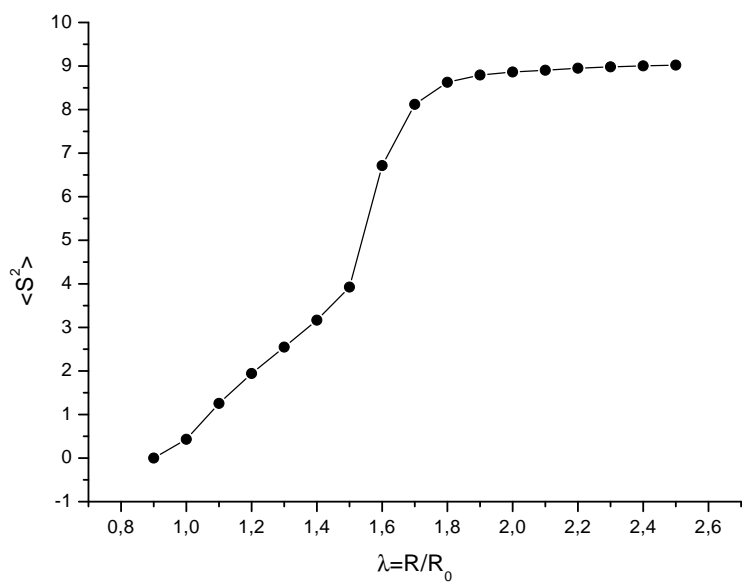}
\caption{Variation of the total spin expectation value $\langle {\bf
   S}^2 \rangle$ for the uniformly scaled benzene ring.
}
\label{fig:A3}
\end{figure}

\begin{figure}[ht]
\centering
\includegraphics[width=0.8\textwidth]{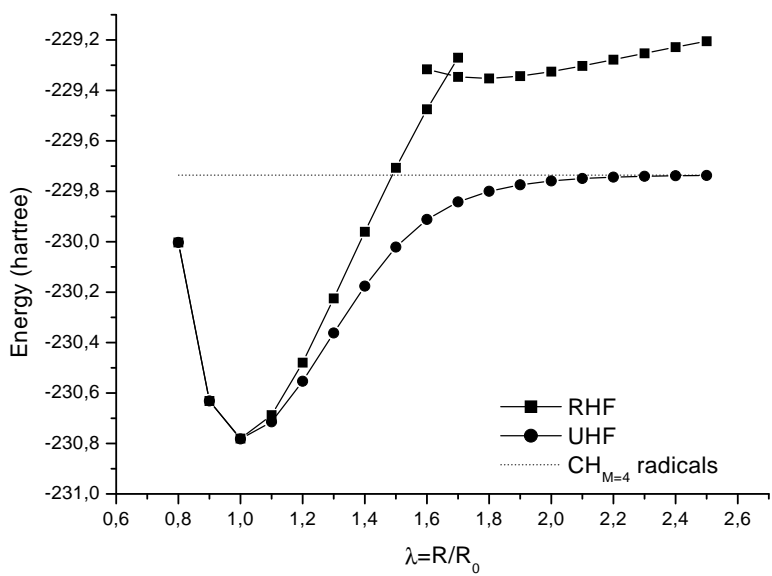}
\caption{Restricted and unrestricted Hartree-Fock potential energy
   curves for \protect\CH.
Carbon ring geometry is scaled with $\lambda$, whereas C-H bond
   lengths are fixed at their equilibrium value.
}
\label{fig:A4}
\end{figure}

\begin{figure}[ht]
\centering
\includegraphics[width=0.8\textwidth]{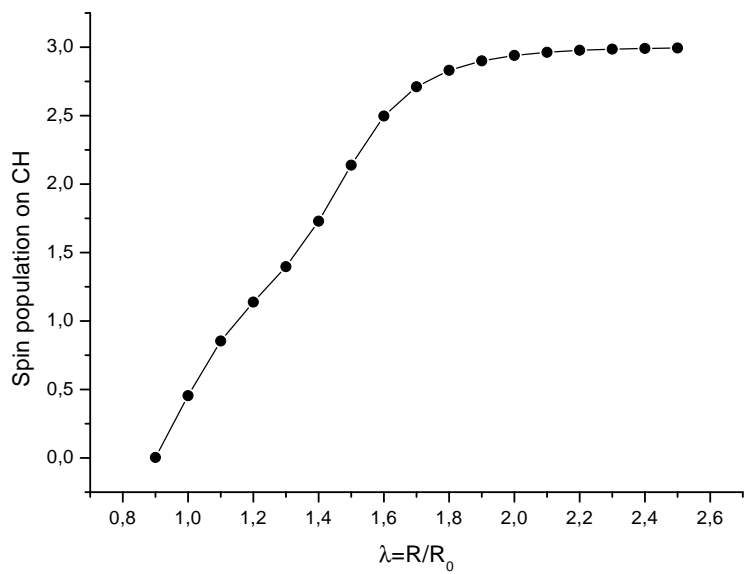}
\caption{Spin population on a CH group in a \protect\CH ring as a function of 
   the ring diameter scaling factor.
}
\label{fig:A5}
\end{figure}

\setcounter{table}{0}

\renewcommand{\thetable}{B.\arabic{table}}

\begin{table}[ht]
\caption{%
Ground state energy and Si--Si and Si--H bond lengths: effect of approximate
   inclusion of electron correlation using two forms of density
   functional theory.
Binding energy $\Delta E$ is related to the dissociation of
   \protect\SiH into 6~SiH radicals.
All energies and lengths are in hartrees and angstroms, respectively.
}
\label{tab:new}
\medskip
\centering
\begin{tabular}{llr@{.}lr@{.}lr@{.}l}
\hline
 & & 
\multicolumn{2}{c}{SiH radical} &
\multicolumn{2}{c}{\SiH} &
\multicolumn{2}{c}{$\Delta E$} \\
\hline
\hline
BLYP & Energy & $-290$ & 00 & $-1740$ & 52 & 0 & 54 \\
     & Geometry & 1 & 545 (Si--H) & 1 & 489 (Si--H)  & \multicolumn{2}{c}{} \\
     &          & \multicolumn{2}{c}{} & 2 & 231 (Si--Si) & \multicolumn{2}{c}{}\\
B3LYP & Energy & $-290$ & 02 & $-1740$ & 66 & 0 & 55 \\
      & Geometry & 1 & 533 (Si--H) & 1 & 481 (Si--H)  & \multicolumn{2}{c}{}\\
      &          & \multicolumn{2}{c}{} & 2 & 215 (Si--Si) & \multicolumn{2}{c}{}\\
UHF  & Energy & $-289$ & 44 & $-1737$ & 04 & 0 & 40 \\
     & Geometry & 1 & 519 (Si--H) & 1 & 473 (Si--H)  & \multicolumn{2}{c}{}\\
     &          & \multicolumn{2}{c}{} & 2 & 214 (Si--Si) & \multicolumn{2}{c}{}\\
\hline
\hline
\end{tabular}
\end{table}

\end{document}